\documentclass[]{spie}  

\usepackage{amsmath,amsfonts,amssymb}
\usepackage{graphicx}
\usepackage[colorlinks=true, allcolors=blue]{hyperref}
\usepackage{enumitem}
\usepackage{siunitx}

\title{The re-flight of the Colorado high-resolution Echelle stellar spectrograph (CHESS): improvements, calibrations, and post-flight results}

\author[a]{Keri Hoadley}
\author[a,b]{Kevin France}
\author[a]{Nicholas Kruczek}
\author[a]{Brian Fleming}
\author[a]{Nicholas Nell}
\author[a]{Robert Kane}
\author[a]{Jack Swanson}
\author[b]{James Green}
\author[b]{Nicholas Erickson}
\author[b]{Jacob Wilson}
\affil[a]{Laboratory for Atmospheric and Space Physics, University of Colorado, UCB 600, Space Science Building (SPSC), 3665 Discovery Drive, Boulder, CO, USA 80309}
\affil[b]{Center for Astrophysics and Space Astronomy, University of Colorado, 389 UCB, Boulder, CO, USA 80309}

\authorinfo{Further author information: (Send correspondence to Keri Hoadley)\\ E-mail: keri.hoadley@colorado.edu, Telephone: +1 303 492 5661}

\begin{document} 
\maketitle

\begin{abstract}
In this proceeding, we describe the scientific motivation and technical development of the \emph{Colorado High-resolution Echelle Stellar Spectrograph} (CHESS), focusing on the hardware advancements and testing supporting the second flight of the payload (CHESS-2).  CHESS is a far ultraviolet (FUV) rocket-borne instrument designed to study the atomic-to-molecular transitions within translucent cloud regions in the interstellar medium (ISM). CHESS is an objective f/12.4 echelle spectrograph with resolving power $>$ 100,000 over the band pass 1000 $-$ 1600 {\AA}.  The spectrograph was designed to employ an R2 echelle grating with ``low" line density. We compare the FUV performance of experimental echelle etching processes (lithographically by LightSmyth, Inc. and etching via electron-beam technology by JPL Microdevices Laboratory) with traditional, mechanically-ruled gratings (Bach Research, Inc. and Richardson Gratings). The cross-dispersing grating, developed and ruled by Horiba Jobin-Yvon, is a holographically-ruled, ``low" line density, powered optic with a toroidal surface curvature. Both gratings were coated with aluminum and lithium fluoride (Al+LiF) at Goddard Space Flight Center (GSFC). Results from final efficiency and reflectivity measurements for the optical components of CHESS-2 are presented. CHESS-2 utilizes a 40mm-diameter cross-strip anode readout microchannel plate (MCP) detector fabricated by Sensor Sciences, Inc., to achieve high spatial resolution with high count rate capabilities (global rates $\sim$ 1 MHz).  We present pre-flight laboratory spectra and calibration results. CHESS-2 launched on 21 February 2016 aboard NASA/CU sounding rocket mission 36.297 UG. We observed the intervening ISM material along the sightline to $\epsilon$ Per and present initial characterization of the column densities, temperature, and kinematics of atomic and molecular species in the observation.   
\end{abstract}


\section{INTRODUCTION}
\label{sec:intro}  

The \emph{Colorado High-resolution Echelle Stellar Spectrograph} (CHESS)\cite{Hoadley+14,France+12,Kane+11,Beasley+10,France+16JAI} is a rocket-borne astronomical instrument that first launched from White Sands Missile Range (WSMR) aboard NASA/CU mission 36.285 UG on 24 May 2014 (CHESS-1). The second flight of the experiment was aboard the NASA/CU mission 36.297 UG on 21 February 2016 (CHESS-2). This paper presents information on the instrument optics, alignment, calibration, and flight results of CHESS-2. This section will cover the instrumental design and scientific objectives of the CHESS experiment.

%
   \begin{figure} [ht]
   \begin{center} 
   \includegraphics[height=4.5cm]
{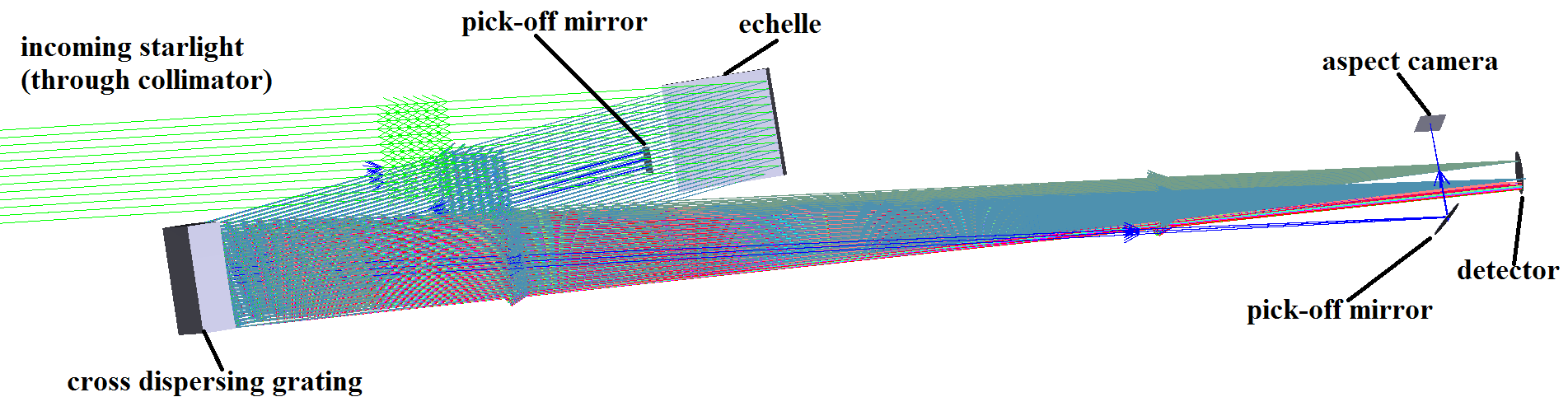}
   \end{center}
   \caption[fig1] 
   { \label{fig:fig1} 
The Zemax ray trace of CHESS, including the secondary aspect camera system.  The mechanical collimator reduces stray light in the line of sight and feeds starlight to the echelle. The echelle disperses UV light into high-dispersion orders, which are focused by the cross disperser onto the detector plane. The different colored lines represent a series of wavelengths across the 1000 $-$ 1600 {\AA} bandpass.}
   \end{figure}

\subsection{Scientific Objectives}

Translucent clouds reside in the transition between the diffuse (traditionally defined as A$_{V}$ $<$ 1) and dense (A$_{V}$ $>$ 3) phases of the interstellar medium (ISM). It is in this regime where the ultraviolet portion of the average interstellar radiation field plays a critical role in the photochemistry of the gas and dust clouds that pervade the Milky Way galaxy. The most powerful technique for probing the chemical structure of translucent clouds is to combine measurements of H$_{2}$ with knowledge of the full carbon inventory (CI, CII, and CO) along a given line of sight. It has been argued that an analysis of the carbon budget should be the defining criterion for translucent clouds, rather than simple measurements of visual extinction\cite{Snow+06}. Moderate resolution 1000 $-$ 1120 {\AA} spectra from FUSE and higher-resolution data from HST/STIS have been used to show that many of these sightlines have CO/H$_{2}$ $>$ 10$^{-6}$ and CO/CI $\sim$ 1, consistent with the existence of translucent material in the framework of current models of photodissociation regions in the ISM\cite{Burgh+07,Burgh+10}. 

\indent The  CHESS experiment is designed to study translucent clouds with its combination of bandpass and spectral resolution.  The 1000 $-$ 1600 {\AA} bandpass contains absorption lines of H$_{2}$ (1000 $-$ 1120 {\AA}), CII (1036 and 1335 {\AA}; however we note that saturation effects can complicate the interpretation of these lines), CI (several between 1103 $-$ 1130, 1261, 1561 {\AA}), and the A $-$ X, B $-$ X, C $-$ X, and E $-$ X bands of CO ($<$ 1510 {\AA}). High resolution (R) $>$ 100,000 is required to resolve the velocity structure of the CI lines and the rotational structure of CO. High-resolution is therefore essential to the accurate determination of the column density of these species\cite{Jenkins+01}. The FUV also provides access to many absorption lines of metals, such as iron, magnesium, silicon, and nickel, allowing for an exploration of the depletion patterns in translucent clouds. CHESS, with its high-resolution and large bandpass, especially including wavelengths shorter than 1150 {\AA}, is well-suited to the study of translucent clouds and will help create an observational base for models of the chemistry and physical conditions in interstellar clouds.

For the second flight of CHESS, we combine our program on the LISM (begun with the NASA/CU mission 36.271 UG/SLICE\cite{France+13a,France+13b,Kane+13} and continued on 36.285 UG/CHESS\cite{Hoadley+14}) with a detailed study of the interstellar material in the line of sight to $\epsilon$ Per (HD 24760). $\epsilon$ Per is a B0.5III star  at d $\approx$ 300 pc with low$-$intermediate reddening (E(B-V) = 0.1; log(H$_2$) $\sim$ 19.5), indicating that the sightline may be sampling cool interstellar material. H$_{2}$, CI, CO, and CII were all detected by Copernicus; however, higher sensitivity and spectral resolution is required for a complete analysis of these types of sightlines\cite{Federman+80}. Observations by Copernicus and IUE have been used to measure the velocity structure along the sightline to $\epsilon$ Per, and have found at least three separate cloud structures described by different kinematic behavior and molecular abundances\cite{Bohlin+83}. Resolving the various molecular clouds on the $\epsilon$ Per sightline is the primary goal of CHESS-2. Overall, the line of sight to $\epsilon$ Per shows typical abundances of molecular material and ionized metal found in translucent clouds, such as H$_{2}$, FeII and MgII\cite{Bohlin+83}, consistent with the sightline towards recent star-forming sites. CHESS-2 provides the first high$-$resolution data of $\epsilon$ Per to observe H$_{2}$, CI, and ionized metal lines simultaneously, allowing us to constrain the metal content, kinematic structure, and photo-dissociation processes of the nearby molecular cloud material around the early B-type star.

%
   \begin{figure} [ht]
   \begin{center}
   \includegraphics[height=12cm]{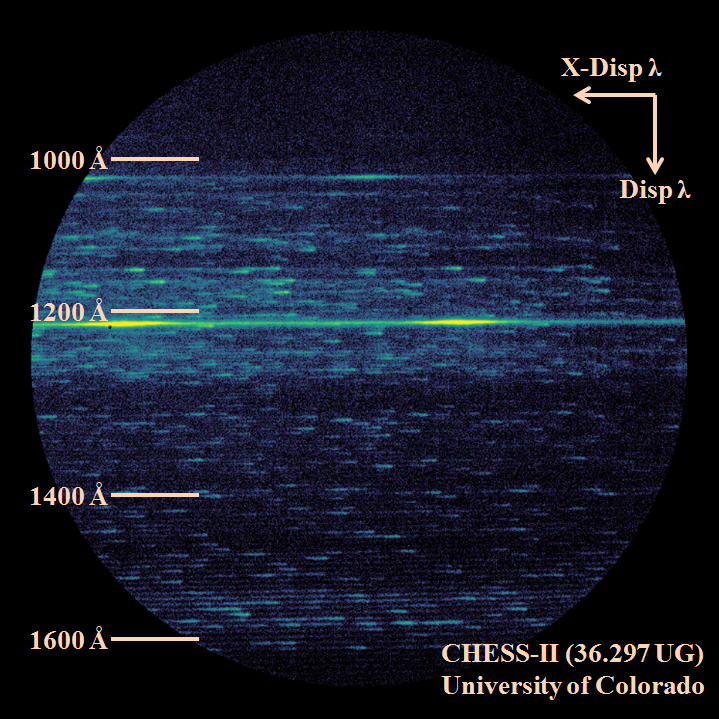}
   \end{center}
   \caption[fig2] 
   { \label{fig:fig2} 
A false-color representation of the full-resolution (8192$\times$8192 digital pixels) laboratory echellogram of CHESS-2 after instrument alignment. The black/blue represents little to no counts in the binned pixel location, while green/yellow represent emission lines from hydrogen (Ly$\alpha$ at 1215.67 \AA) and molecular hydrogen. Overlaid are arrows showing the direction of dispersion from the echelle and cross disperser, and the tick marks show the approximate location of different spectral regions. The final laboratory calibration image contains over 73 million photon counts.}
   \end{figure}

\subsection{Instrument Design}

CHESS is an objective f/12.4 echelle spectrograph. The instrument design\cite{Beasley+10} included the development of two novel grating technologies and flight-testing of a cross-strip anode microchannel plate (MCP) detector. The high-resolution instrument is capable of achieving resolving powers $\ge$ 100,000 $\lambda$/$\Delta \lambda$ across a bandpass of 1000 $-$ 1600 {\AA}\cite{France+16JAI}. The operating principle of the instrument is as follows:

\begin{itemize}

\item A mechanical collimator, consisting of an array of 10.74 mm $\times$ 10.74 mm $\times$ 1000 mm anodized aluminum tubes, provides CHESS with a total collecting area of 40 cm$^{2}$, a field of view (FOV) of 0.67$^{\circ}$, and allows on-axis stellar light through to the spectrograph.

\item A square echelle grating (ruled area: 100 mm $\times$ 100 mm), with a designed groove density of 69 grooves/mm and angle of incidence (AOI) of 67$^{\circ}$, intercepts and disperses the FUV stellar light into higher diffraction terms (m = 266 $-$ 166). The custom echelle was used to research new etching technologies, namely the electron-beam (e-beam) etching process. The results of the JPL fabrication effort are discussed in Section~\ref{sec:sec2.1}.

\item Instead of using an off-axis parabolic cross disperser\cite{Jenkins+88}, CHESS employs a holographically-ruled cross dispersing grating with a toroidal surface figure and ion-etched grooves, maximizing first-order efficiency\cite{France+16JAI}. The cross disperser is ruled over a square area (100 mm $\times$ 100 mm) with a groove density of 351 grooves/mm and has a surface radius of curvature (RC) = 2500.25 mm and a rotation curvature ($\rho$) = 2467.96 mm. The grating spectrally disperses the echelle orders and corrects for grating aberrations\cite{Thomas+03}.

\item The cross-strip MCP detector\cite{Vallerga+10,Siegmund+09} is circular in format, 40 mm in diameter, and capable of total global count rates $\sim$ 10$^{6}$ counts/second. The cross-strip anode allows for high resolution imaging, with the location of a photoelectron cloud determined by the centroid of current read out from five anode ``fingers" along the x and y axes.

\end{itemize}

The CHESS experiment also includes an optical system (aspect camera), used solely to align the spectrograph to the stellar target during calibrations and flight. The aspect camera system uses the positions of the instrument gratings to direct zeroth-order light to a stand-alone camera. Fig.~\ref{fig:fig1} shows a Zemax ray trace of collimated light through the spectrograph, and Fig.~\ref{fig:fig2} displays an example of a laboratory spectrum taken with the fully integrated and aligned CHESS instrument.

\subsection{Mechanical Spectrograph Structure}
\label{sec:sec1.3}

\begin{figure} [ht]
   \begin{center}
   \includegraphics[height=6cm]{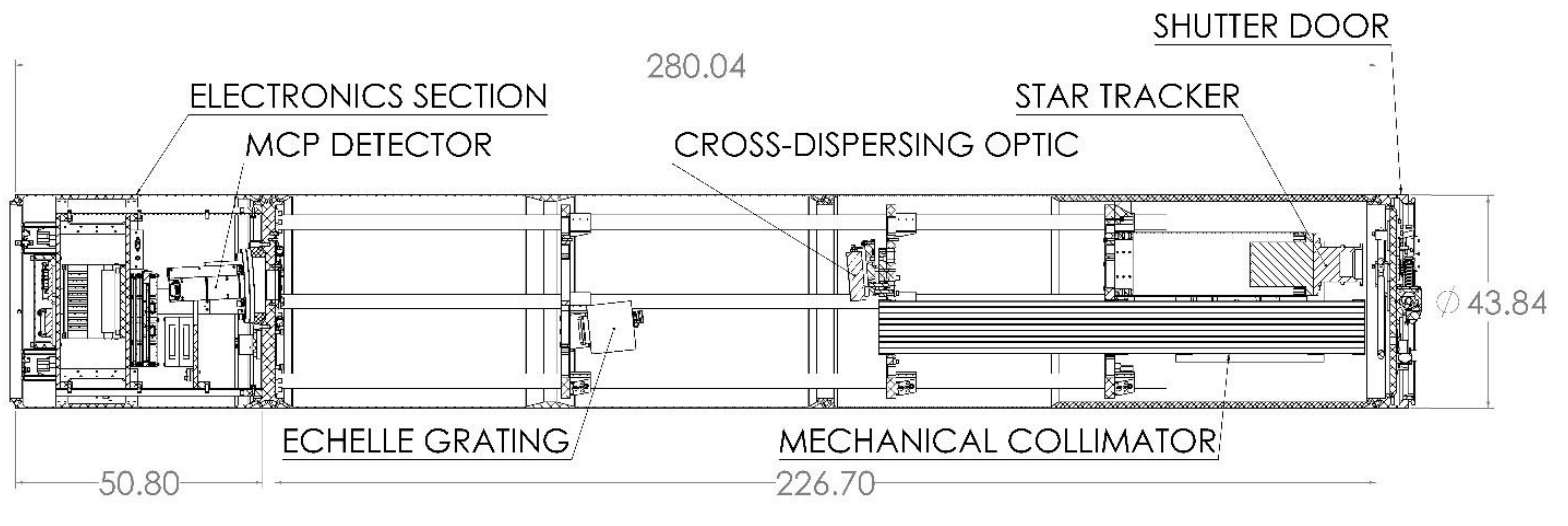}
   \end{center}
   \caption[fig3] 
   { \label{fig:fig3} 
A SolidWorks rendering (all stated quantities in units of centimeters) of the spectrograph and electronics sections of CHESS (skins excluded). Labeled are relevant spectrograph structures and optical components referenced in Section~\ref{sec:sec1.3}.}
   \end{figure}

The opto-mechanical structure of CHESS-2 is identical to that presented in Ref.~\citenum{Hoadley+14}. Below, we outline information and specifications about the structure and optical mount designs for CHESS. A SolidWorks rendering of the spectrograph and electronics sections of CHESS is provided in Fig.~\ref{fig:fig3}.

%

CHESS is an aft-looking payload that uses 17.26 inch-diameter rocket skins and is split into two sections: a vacuum (spectrograph) section and non-vacuum (electronics) section. The two sections are separated by a hermetic bulkhead. The overall length of the payload is 226.70 cm from mating surface to mating surface and the weight of the payload is 364 lbs. The shutter door is the only moving component during flight in the experiment section. The electronics section is 50.80 cm long with one umbilical pocket for an RJ45 Ethernet port. The detector is mounted with a hermetic seal on the electronics side of the vacuum bulkhead, facing into the spectrograph section.

The vacuum section uses two 113.36 cm long rocket skins with hermetic joints. The only mechanical component on CHESS (other than the NASA Sounding Rockets Operations Contact (NSROC)-supplied shutter door) is a manual butterfly valve attached along the 180$^{\circ}$ line on the aft skin. This allows for the evacuation of the experiment throughout development, integration and pre-flight activities, safeguarding the sensitive optical coatings. A carbon-fiber space frame is attached to the aft side of the hermetic bulkhead and suspends the aspect camera, mechanical collimator, echelle grating and cross-disperser in place. The space frame is comprised of three aluminum disks attached to five 2.54 cm diameter x 182.88 cm long carbon fiber tubes.

The collimator is a set of 10.74 mm x 10.74 mm x 1000 mm long black-anodized aluminum square tubes bonded together. An aluminum mounting flange is bonded at the center of the assembly and secured to the aft-most disk of the space frame.  

The echelle grating in CHESS-2 is a rectangular block made of Zerodur (110 mm x 110 mm x 16 mm thick). To provide support during launch, the grating is bonded to a mount via three Invar brackets. Three titanium flexures secure the Invar brackets to an aluminum mount, which is set to the desired AOI and is attached to the forward most disk of the space frame.  The flexures prevent stress transfer into the optic due to a coefficient of thermal expansion (CTE) mismatch between mounting components. 

The cross disperser is a block of fused silica (100 mm x 100 mm x 30 mm thick) with a toroidal surface.  Three Invar pads are bonded to the neutral plane of the optic, nearly 120$^{\circ}$ apart, and are connected to three titanium flexures.  The flexures are attached to an aluminum mounting plate and affixed to the middle disk of the space frame.  Again, the flexures prevent the surface of the optic from warping due to stress transfer from a CTE mismatch in mounting components. 

%
   \begin{figure} [ht]
   \begin{center}
   \includegraphics[height=8cm]{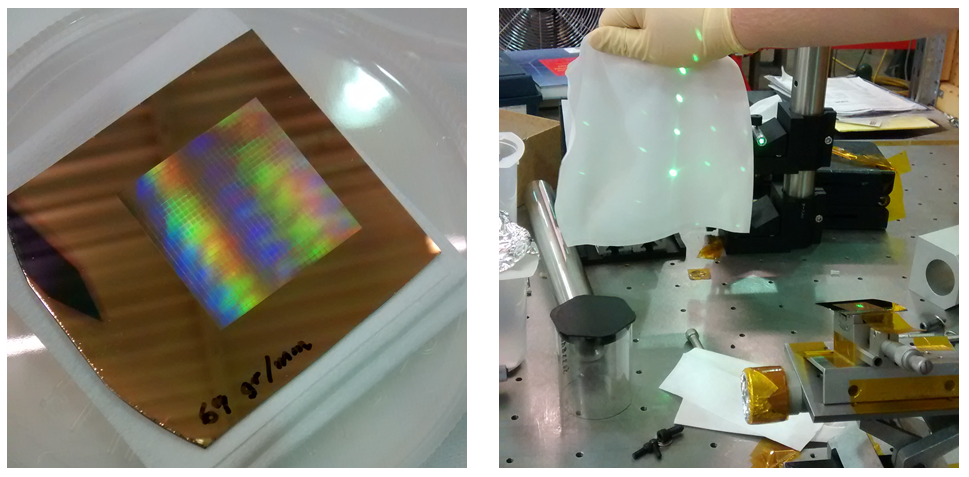}
   \end{center}
   \caption[fig4] 
   { \label{fig:fig4} 
\emph{Left}: An image of one of the sample e-beam etched echelle gratings (fabricated by JPL Microdevices Laboratory) tested at CU. The echelle was etched into a PMGI photo-resist overlaid on a silicon substrate and coated with gold for reflectivity measurements in the UV. The dispersion direction of the echelle can be seen by the optical colors on the grating in the photo. There are small squares over the entire grated area of the echelle, due to the periodicity in the etching procedure.
\emph{Right}: An image showing the optical diffraction off of one of the experimental echelle samples from JPL. The JPL e-beam echelles showed secondary diffraction patterns parallel to the primary diffraction axis (``ghosts"), which we believe are a consequence of the periodicity in the etching process. In the FUV, ghost orders contained as much as 60\% of the total counts as their primary order counterparts. In practice, this would have lead to at least 3 overlapping echellograms (primary + two sets of ghost orders) in CHESS, making the data complicated to decipher.}
   \end{figure}

\section{OPTICAL PERFORMANCE}
\label{sec:optics}

\subsection{Echelle Gratings}
\label{sec:sec2.1}

The CHESS echelle was designed to be a 100 mm x 100 mm x 0.7 mm silicon wafer with a groove density of 69 grooves/mm and AOI = 67$^{\circ}$.  The CHESS-1 echelle was the final result of a research and development lithography-etching project undertaken by LightSmyth, Inc., meant to suppress scattered light in the FUV, making the grating more efficient in the peak echelle orders. Unfortunately, the etching process was difficult to perfect, and CHESS-1 flew an echelle grating with peak order efficiencies $\sim$ 1 $-$ 6\% across the bandpass\cite{Hoadley+14}.

To improve the sensitivity of the CHESS instrument for its second flight, we worked in collaboration with the JPL Microdevices Laboratory on an experimental electron-beam (e-beam) etching process to fabricate an echelle grating with low-scatter and high order efficiency. Theoretically, groove efficiency expectations at the peak order in the FUV were around 85\%. Practically, we specified a groove efficiency of 60\% for JPL to fabricate to demonstrate the e-beam technology for future UV observatories, while the CHESS-2 primary scientific objectives required a minimum groove efficiency of 20\% across the bandpass 1000 $-$ 1600 {\AA}. The e-beam process was noted to have better precision and groove etching control than either the mechanical-ruling or lithographic-etching processes\cite{Wilson+03}. An image of one of the samples from JPL is shown in Figure~\ref{fig:fig4}. The first sample was etched directly onto the silicon substrate, which proved difficult for JPL to control and resulted in a jagged saw-tooth grating wall. All other grating profiles were fabricated into a thin layer of polymethylglutarimide (PMGI), which was deposited onto the silicon substrate. PMGI is a photo-resist that is more malleable than silicon, making the etching process more precise and easier to control. To measure the order efficiency of the echelles in the FUV, each PMGI grating sample was gold-coated.

%
   \begin{figure} [ht]
   \begin{center}
   \includegraphics[height=6cm]{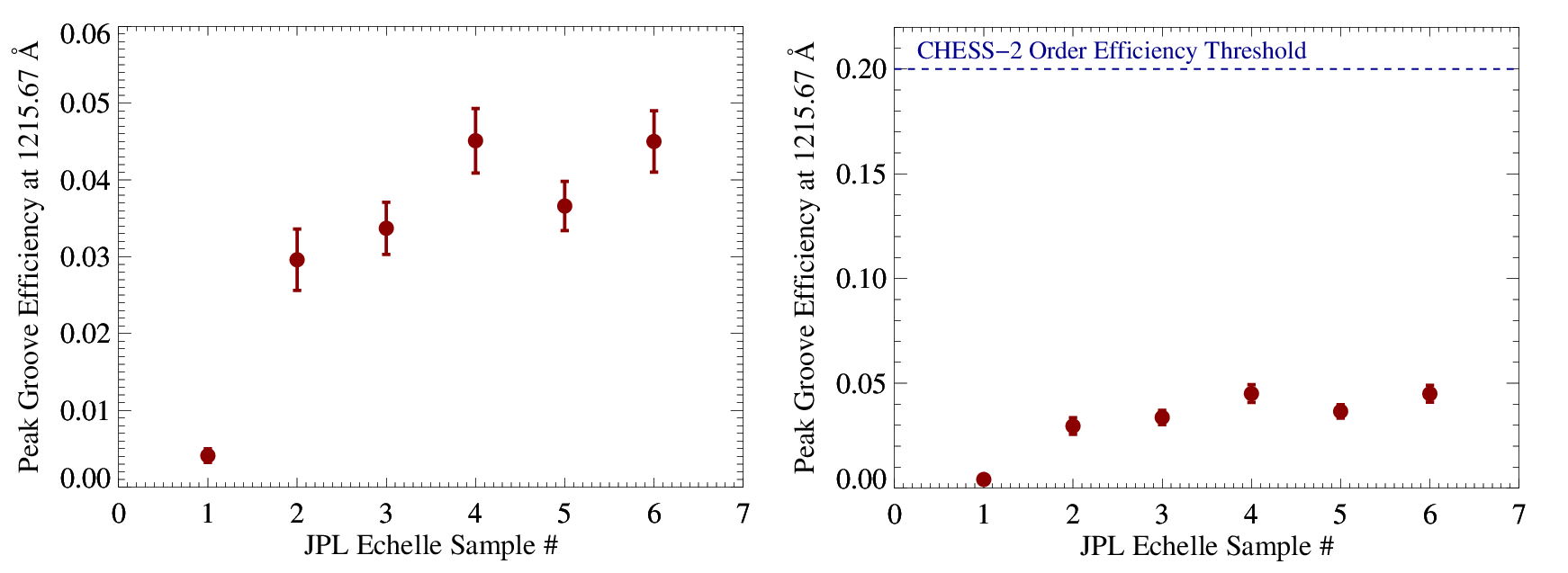}
   \end{center}
   \caption[fig5] 
   { \label{fig:fig5} 
A comparison of the peak order groove efficiency (1.0 = 100\% efficient) of each JPL e-beam echelle sample at HI-Ly$\alpha$ (1215.67 {\AA}). The first sample (\#1) was etched directly into the silicon substrate, which created jagged groove profiles and caused much of the light to be lost to scatter. Samples \#2 - \#6 were etched into PMGI, which was overlaid on the silicon substrate and coated with gold. The best performance achieved by the samples was $\sim$4.5\% peak order efficiency (Samples \#4 and \#6), which incidentally were the samples etched with 100 grooves/mm instead of the 69 grooves/mm. We include the CHESS-2 threshold order efficiency (\emph{right}). This threshold represents the minimum order efficiency required of the echelle grating to observe the CHESS-2 target ($\epsilon$ Per) at a S/N $\sim$ 20 over the 250-second exposure time.}
   \end{figure}

Figure~\ref{fig:fig5} shows the groove efficiency of each CHESS echelle grating sample provided by JPL between February $-$ June 2015. All JPL e-beam gratings failed to meet the specified target groove efficiency for the project. We reiterate that the original groove efficiency specification given to JPL was 60\%, while the science goals of the CHESS-2 mission required a groove efficiency of at least 20\% from 1000 $-$ 1600 {\AA}. The most efficient gratings (Samples \#4 and \#6) were only 4.5\% efficient at Ly$\alpha$, which was comparable to the efficiency of the LightSmyth, Inc. echelle flown in CHESS-1\cite{Hoadley+14}. We also saw noticeable ghosting effects $-$  secondary diffraction patterns parallel to the primary diffraction axis $-$ from every echelle sample fabricated with the e-beam technique, both in the optical and in the FUV.
We suspect this was due to the periodicity in the e-beam process, which left a grid pattern of small squares over the entire grated surface area. Figure~\ref{fig:fig4} shows an image of the grid-like pattern and the resulting ghosting off the grating with a 532 nm laser. We note that the R\&D project had difficulty meeting the CHESS echelle parameters throughout the fabrication process. We list the deviations from the optimal AOI (67$^{\circ}$) and groove density (69 grooves/mm) of each sample in Table~\ref{tab:tab1}. 

%
   \begin{figure} [ht]
   \begin{center}
   \includegraphics[height=9cm]{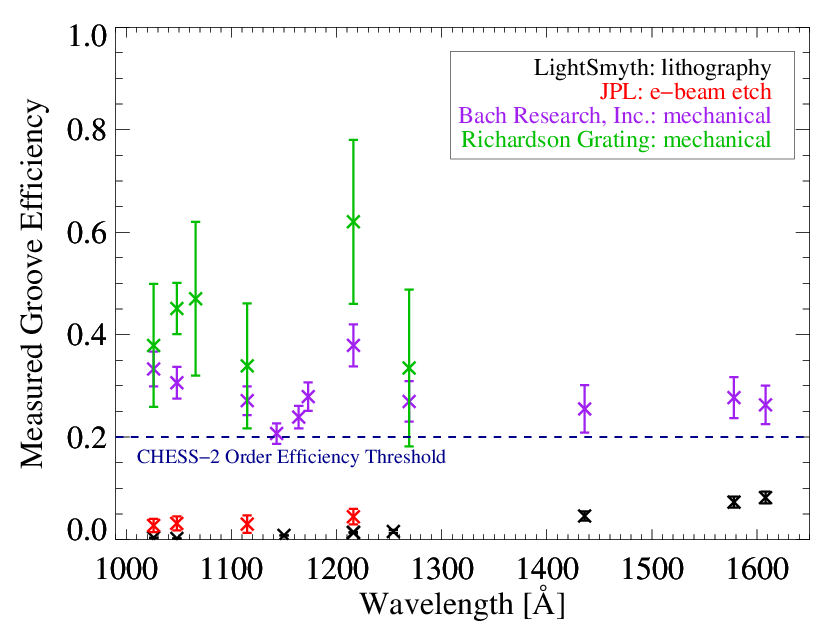}
   \end{center}
   \caption[fig6] 
   { \label{fig:fig6} 
A comparison of echelle gratings tested for use in the CHESS instrument. We include the best-performing echelle gratings from the lithography etching R\&D project undertaken by LightSmyth, Inc. (flown on CHESS-1, 36.285 UG), the e-beam samples fabricated by JPL, and two mechanically-ruled replica gratings from Bach Research, Inc. and Richardson Gratings, respectively. Both mechanically-ruled gratings out-performed the R\&D echelles and met the CHESS minimum order efficiency threshold.}
   \end{figure}

Due to ongoing challenges with the e-beam process, we opted to explore traditional, mechanically-ruled echelle gratings, which can have higher efficiency but historically display higher inter-order scatter\cite{Landsman+97}. We tested several echelle gratings from two manufactures with histories of providing UV space missions with gratings: Bach Research Inc. (formerly Hyperfine Inc. $-$ Boulder, CO) and Richardson Gratings (formerly Milton-Roy $-$ Rochester, NY). Figure~\ref{fig:fig6} shows a comparison of the best-performing gratings by each manufacturer (LightSmyth, Inc.; JPL; Bach Research, Inc.; and Richardson Gratings). The mechanically-ruled echelle gratings outperformed all of the experimental gratings. Given time constraints on the delivery of the grating to meet the launch date of CHESS-2 and their vicinity to CU, we chose to fly the Bach echelle grating for CHESS-2. We note that we have ordered and currently have in-house a Richardson echelle grating ($\alpha$ = 63$^{\circ}$, g = 87 grooves/mm), which both outperformed all echelle gratings tested at CU in the FUV and has the closest matching echelle solutions to those designed for the CHESS instrument (m = 266 $-$ 166 for $\lambda$ = 1000 $-$ 1600 {\AA}). It will be flown on CHESS-3 (June 2017) and CHESS-4 (early 2018).

%
%
\begin{table}[ht]
\caption{Comparison of Echelle Parameters Tested for CHESS-2 and fabricated gratings.} 
\label{tab:tab1}
\begin{center}       
\begin{tabular}{|l|c|c|c|} 
\hline
\rule[-1ex]{0pt}{3.5ex}  Grating Label & Angle of Incidence ($^{\circ}$) & Groove Density (grooves/mm) & Ly$\alpha$ Groove Efficiency \\
\hline
\rule[-1ex]{0pt}{3.5ex}  \textbf{CHESS Echelle, Designed} & \textbf{67.0} & \textbf{69.0} & \textbf{60.0 \%}  \\
\rule[-1ex]{0pt}{3.5ex}  LightSmyth, CHESS-1 Flight & 73.0 & 71.7 & 1.5 \%  \\
\rule[-1ex]{0pt}{3.5ex}  JPL, Echelle Sample \#1 & 57.7 & 83.3 & 0.4 \% \\
\rule[-1ex]{0pt}{3.5ex}  JPL, Echelle Sample \#2 & $\sim$67 & $\sim$69 & 2.9 \% \\
\rule[-1ex]{0pt}{3.5ex}  JPL, Echelle Sample \#3 & 68.6 & 70.0 & 3.4 \%  \\
\rule[-1ex]{0pt}{3.5ex}  JPL, Echelle Sample \#4$^{\star}$ & 72.1 & 100.0 & 4.5 \%  \\
\rule[-1ex]{0pt}{3.5ex}  JPL, Echelle Sample \#5 & 66.7 & 69.0 & 3.7 \% \\
\rule[-1ex]{0pt}{3.5ex}  JPL, Echelle Sample \#6$^{\star}$ & 65.5 & 100.0 & 4.5 \% \\
\rule[-1ex]{0pt}{3.5ex} Bach Echelle$^{\star}$ & 64.3 & 53.85 & 11.9 \% + 8.4 \% \\
\rule[-1ex]{0pt}{3.5ex}  Richardson, Echelle \#1$^{\star}$ & 64.8 & 79.0 & 30.0 \% \\
\rule[-1ex]{0pt}{3.5ex}  Richardson, Echelle \#2$^{\star}$ & 63.0 & 87.0 & 62.0 \% \\
\hline 

\multicolumn{4}{l}{$^{\star}$ These echelle gratings were not specified to achieve the designed groove density of the CHESS instrument.}
\end{tabular}
\end{center}
\end{table}

\subsection{Cross Disperser Grating}

The CHESS cross disperser grating is a 100 mm $\times$ 100 mm $\times$ 30 mm fused silica optic with a toroidal surface profile. The toroidal surface shape separates the foci of the spatial and sagittal axes of the dispersed light. The optic first focuses light spatially onto the detector, then spectrally behind the detector, ensuring no foci at the locations of either the ion repeller or quantum efficiency (QE) grids. The cross dispersing optic is a novel type of imaging grating that represents a new family of holographic solutions and was fabricated by Horiba Jobin-Yvon (JY). The line densities are low (351 lines per mm, difficult to achieve with the ion-etching process), and the holographic solution allows for more degrees of freedom than were previously available with off-axis parabolic cross dispersing optics. The holographic ruling corrects for aberrations that otherwise could not be corrected via mechanical ruling. The grating is developed under the formalism of toroidal variable line spacing gratings\cite{Thomas+03} and corresponds to a holographic grating produced with an aberrated wavefront via deformable mirror technology. This results in a radial change in groove density and a traditional surface of concentric hyperboloids from holography, like those used in ISIS\cite{Beasley+04} and HST/COS\cite{Green+03}.

%
   \begin{figure} [ht]
   \begin{center}
   \includegraphics[height=7.5cm]{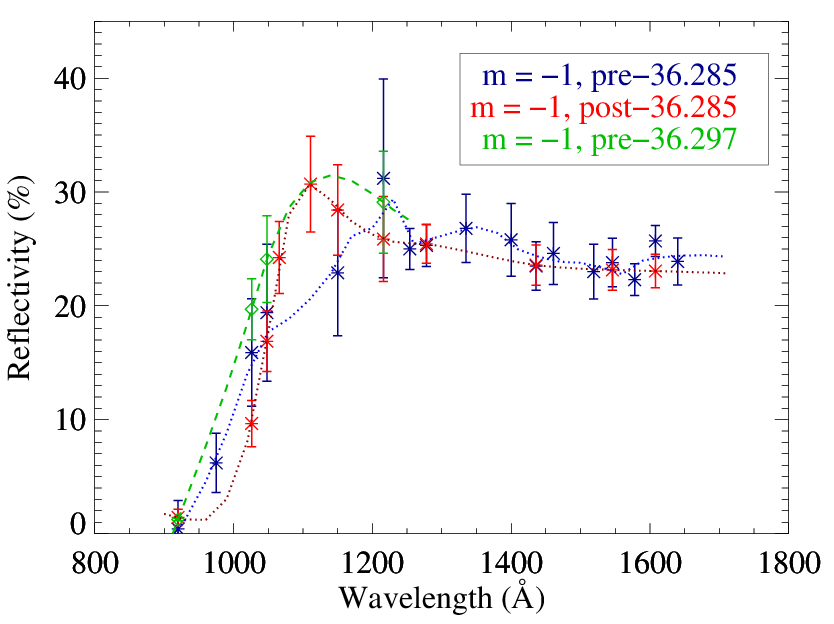}
   \end{center}
   \caption[fig7] 
   { \label{fig:fig7} 
Measured reflectivity (order efficiency $\times$ reflectivity of Al+LiF) of the cross dispersing grating in CHESS over time, overplotted with simple spline curves to show the resemblance of each trial. We focus on the reflectivity of the m = -1 order, which is the dispersion order used in the CHESS instrument. Because LiF can show efficiency degradations when not stored properly, we measure how the order reflectivity changes between CHESS-1 and CHESS-2 without re-coating the optic. No significant degradation of the coating has been measured between the first assembly of CHESS (November 2013) and the build-up of CHESS-2 (November 2015).}
   \end{figure}

The cross disperser was delivered in the summer of 2012, and order efficiencies around both the m = +1 and m = -1 orders were measured to be between 20\% $-$ 45\% in the FUV (900 $-$ 1700 {\AA}) before and after the Al+LiF optical coating. Figure~\ref{fig:fig7} shows the reflectivity (order efficiency $\times$ reflectivity of Al+LiF) of the cross dispersing optic for order m = -1, which is the dispersion order used in the CHESS instrument, for pre-36.285 field operations, post-36.285 launch, and pre-36.297 field operations. Overall, the performance of the cross disperser exceeded our initial expectations, with reflectivity $\sim$ 30\% at Ly$\alpha$. The cross disperser is effective at dispersing most of the on-axis light into the m = $\pm$ 1 orders and suppressing the m = 0 order because of the characteristic sinusoidal groove profiles created via the ion-etching procedure at JY. Additionally, at optical wavelengths, the reflectivity of the m = 0 order becomes comparable to the m = $\pm$ 1 orders. This allowed us to build a secondary camera system to track the movements of our optical axis and target acquisition during flight.

\subsection{Cross-Strip Anode Microchannel Plate Detector}

\begin{figure} [ht]
   \begin{center}
   \includegraphics[height=6.5cm]{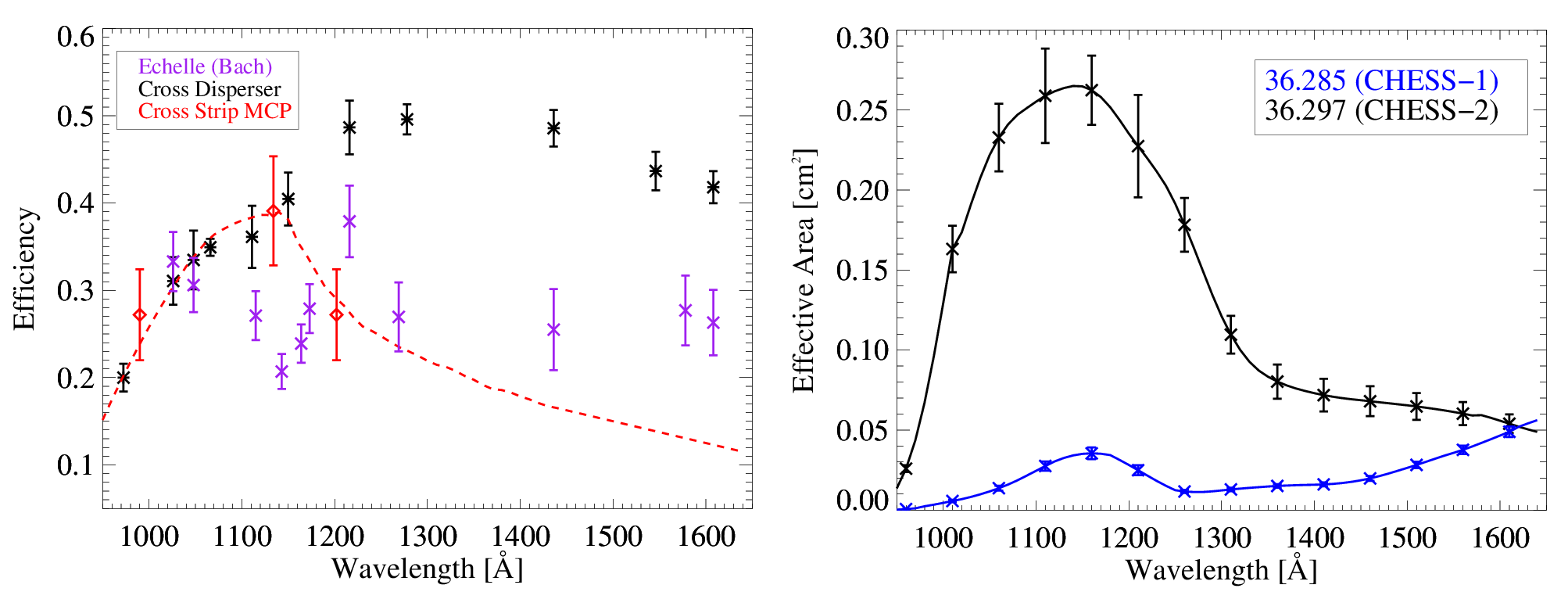}
   \end{center}
   \caption[fig8] 
   { \label{fig:fig8} 
\emph{Left:} Performance (for each grating: peak order efficiency, and for the detector: detector quantum efficiency) of all optical components of CHESS-2. \emph{Right:} The CHESS-2 effective area, including throughput loss from baffling, compared to the effective area of CHESS-1. After 36.285, we sent the cross-strip anode MCP back to Sensor Sciences, Inc.  to replace the CsI photocathode, which had begun to crystallize during field operations of CHESS-1. The total effective area of CHESS-2 is about an order of magnitude larger than that of CHESS-1, deriving mainly from the large gain in echelle order efficiency across the CHESS bandpass\cite{Hoadley+14} (see Figure~\ref{fig:fig6} for a comparison of the echelle performance).
}
   \end{figure}

The cross-strip MCP detector was built and optimized to meet the CHESS spectral resolution specifications at Sensor Sciences\cite{Vallerga+10,Siegmund+09}. The detector has a circular format and a diameter of 40 mm. The microchannel plates are lead silicate glass, containing an array of 10-micron diameter channels. They are coated with an opaque cesium iodide (CsI) photocathode, which provides QE = 15 $-$ 35\% at FUV wavelengths. When UV photons strike the photocathode to release photoelectrons, the photoelectrons are accelerated down the channels by an applied high voltage ($\sim$ 3100 V). Along the way, they collide with the walls of the channels, which produces a large gain over the initial single photoelectron. There are two MCPs arranged in a ``chevron" configuration. During flight, the detector achieved spatial resolution of 25 $\mu$m over an 8k x 8k pixel format. The QE estimate across the CHESS bandpass, measured by Sensor Sciences, is plotted in Figure~\ref{fig:fig8} against the efficiency measurements of the flight echelle and cross disperser for CHESS-2.

%

The 2015 NASA Cosmic Origins Program Annual Technology Report emphasized that the technology readiness level (TRL) for large format, high count rate, and high QE MCP detectors needs to improve for future UV space missions. One of the goals of the CHESS instrument is to demonstrate the flight performance of the cross strip anode design to raise the TRL level to 6, which was achieved on 36.285. The cross-strip anode MCP detector was re-flown and performed reliably once more on 36.297, handling count rates of 25,000 photons/second for the entire exposure of CHESS-2. However, in a laboratory setting, we have been able to demonstrate count rates of $\gtrsim$ 150,000 photons/second, which the MCP handled smoothly.

\section{Instrument alignments and calibrations}
\label{sec:cals}

\begin{figure} [ht]
   \begin{center}
   \includegraphics[height=10.5cm]{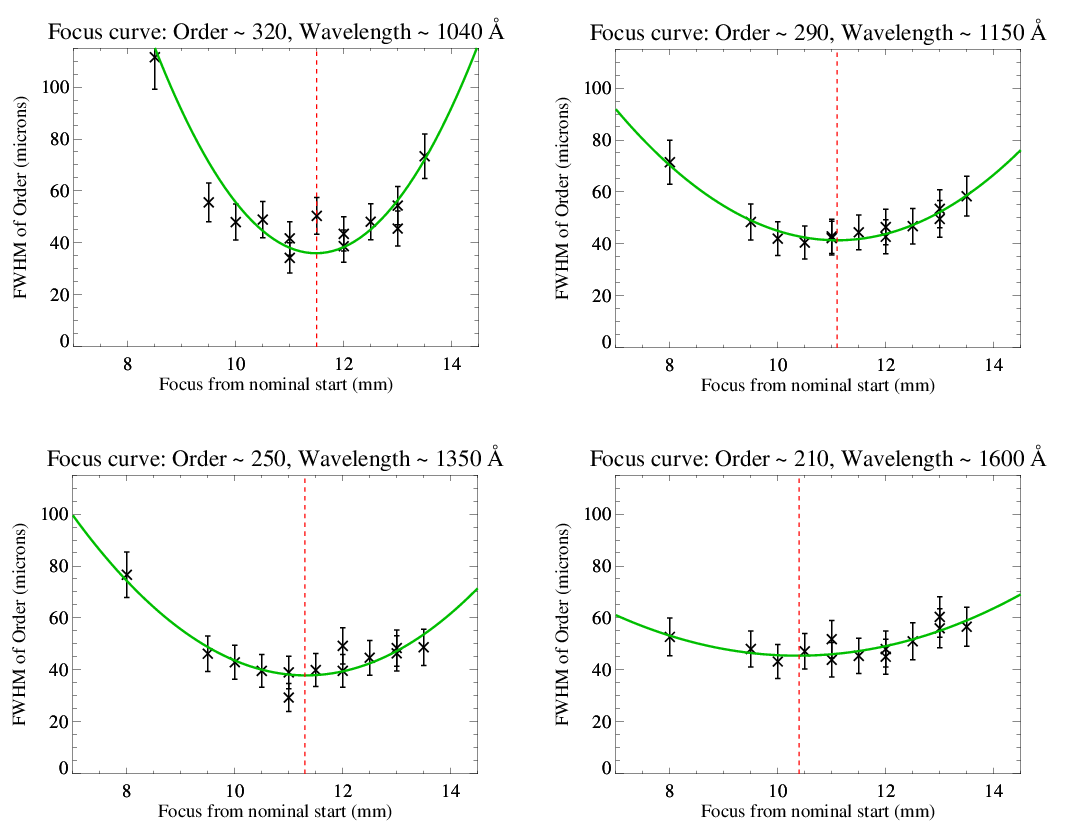}
   \end{center}
   \caption[fig9] 
   { \label{fig:fig9} 
Focus curves for CHESS-2 echellogram orders with $\lambda$ $=$ 1040 {\AA}, 1150 {\AA}, 1350 {\AA}, and 1600 {\AA}. Full width at half maximum (FWHM) measurements were taken as the overall order width, not the spectral width of specific emission features within the order. We focused the echellogram closer to the minima of the 1040 {\AA} and 1150 {\AA} orders, at 11.2 mm from the original starting position of the cross disperser during FUV alignments, because the separation of orders with wavelengths $\lambda$ $<$ 1100 {\AA} was critical in the final data product. 
}
   \end{figure}

The alignment process for CHESS has been described in great detail by Ref.~\citenum{Hoadley+14}; please refer to this document for specifics and pictures of the process. Instead of reiterating the entire procedure in detail, we list the steps taken to align CHESS-2 and focus the echellogram for flight:
\begin{itemize}
	\item The CHESS gratings were specifically designed to have grating parameters to allow for optical wavelength solutions. Grating solutions were modeled in Zemax for red (632 nm), green (532 nm), and violet (405 nm) wavelengths. We utilize this tool to first align the echelle to the cross disperser, and then align the cross disperser to the detector.
    \item For FUV calibrations, the instrument is aligned to an external vacuum system to feed the spectrograph with UV light. We used the aspect camera system to align the instrument to the vacuum chamber. A small, square mirror was set up along one side of the echelle grating to intercept light, which was directed to the center of the cross disperser. Zeroth-order optical light off the cross disperser intercepted another small, flat mirror at the detector bulkhead, which steered the light to the aspect camera. During alignments, white light was tracked with the aspect camera until the proper alignment position was found.
    \item Linear vacuum actuators were positioned behind the cross dispersing optic to control the tip, tilt, and optical axis motions of the grating. Using these stages, we centered the echellogram on the detector and focused the spectrum using small actuator steps. The final focused position of the cross disperser was found to be 11.2 $\pm$ 0.3 mm from the starting position of the cross disperser. Figure~\ref{fig:fig9} shows the focus curves for four different wavelengths that are representative of different portions of the echellogram. 
\end{itemize}

%

%
   \begin{figure} [ht]
   \begin{center}
   \includegraphics[height=8cm]{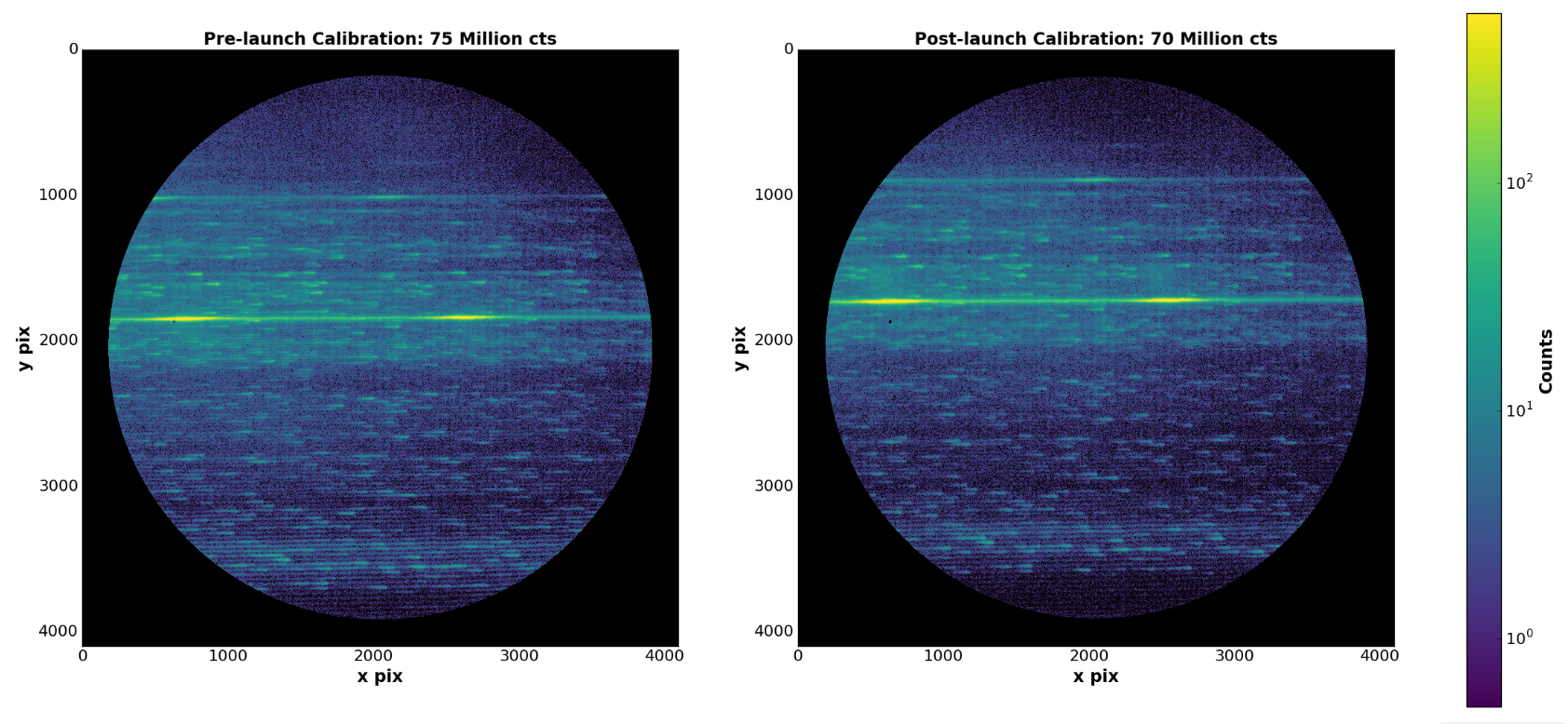}
   \end{center}
   \caption[fig10] 
   { \label{fig:fig10} 
Raw images (with edge effects trimmed out) of the CHESS-2 echellograms from pre-flight (December 2015) and post-flight calibrations (March 2016) using an arc lamp flowing H/Ar gas. The brightest feature in both images is HI-Ly$\alpha$ (1215.67 {\AA}), which appears in two adjacent orders in the echellogram. The other broad feature(s) are HI-Ly$\beta$ (1025.72 {\AA}), about 1/4 of the way from the top of the image, and HI-Ly$\gamma$ (97.25 {\AA}), barely visible above the Ly$\beta$ features. Both images are scaled the same. Between field operations, launch, and returning to CU, the instrument echellogram shifted slightly to the left and upward in the images shown, corresponding to total shifts $<$ 5$'$.
}
   \end{figure}

After final alignment and focus positions of the echelle and cross disperser were determined, long exposures with a 65/35\% hydrogen/argon (H/Ar) gas mixture fed through a hollow-cathode (``arc") lamp were taken for a complete sampling of H and H$_2$ emission lines in the CHESS bandpass. Pre- and post-launch deep spectra with the H/Ar lamp are taken to characterize the one-dimensional (1D) extracted spectrum, define the wavelength solutions of the instrument, and determine the line spread functions (LSFs) across the bandpass. 

Figure~\ref{fig:fig10} shows the echellogram of CHESS-2 for pre- and post-launch calibrations. Both echellograms are co-additions of multiple exposures taken under vacuum to accumulate more than 60 million photon counts for a complete sampling of H and H$_2$ emission lines. Each exposure was defined by how long we could run the full instrument configuration in vacuum without over-heating the electronics section, which usually lasted around 30 minutes. Each exposure collected was typically between 10 - 20 million photon counts.

Extracting the 1D spectra from the echellogram was accomplished in several steps. First, the echellogram had to be rotated very slightly ($\theta$ $<$ 1$^{\circ}$) to successfully extract the light in each order without contamination from light in adjacent orders. The location of each order was then determined by summing (collapsing) all photon counts along the x-axis, which added all the light in each order together and created peaks where orders were present and troughs at inter-order pixels. This exercise also determined the width of each order, which ranged from 4 $-$ 12 pixels wide in the CHESS-2 echellogram.

%
   \begin{figure} [ht]
   \begin{center}
   \includegraphics[height=9.5cm]{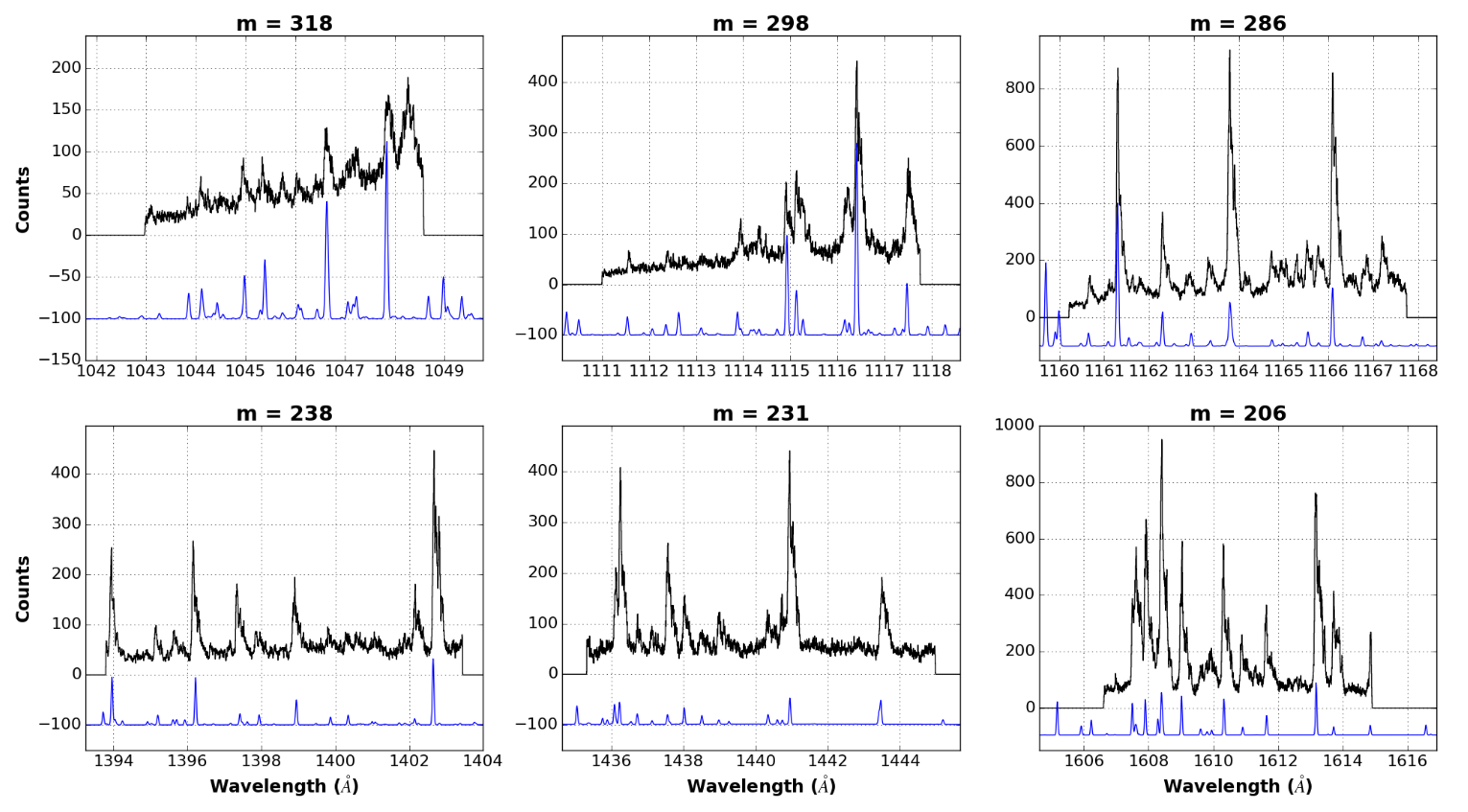}
   \end{center}
   \caption[fig11] 
   { \label{fig:fig11} 
The extracted 1D spectra of six orders from the pre-flight calibration echellogram of CHESS-2. Black represents the CHESS-2 extracted spectrum over the order extent. The blue lines are modeled H$_2$ emission features, using estimated physical parameters of the conditions within the arc lamp, including the column density of H$_2$ molecules (N(H$_{2}$) $\sim$ 10$^{19}$ cm$^{-2}$), effective temperature (T$_{eff}$ = 800 K), and electron energy (E$_{elec}$ = 50 eV). 
}
   \end{figure}

Once pixel locations and order widths were extracted, we collapsed a given order along the y-axis, creating the 1D spectrum of each order. We used the composition of air through the arc lamp to map out well-known atomic lines and their corresponding orders. Once wavelengths and order locations were known for these emission lines, we used the H/Ar arc lamp echellogram, in conjunction with modeled H$_2$ fluorescent lines, to extrapolate the pixel-to-wavelength conversion for CHESS-2 over the entire FUV bandpass; examples of the pixel-to-wavelength extractions determined from H$_2$ emission features are shown in Figure~\ref{fig:fig11}. 

\begin{figure} [ht]
   \begin{center}
   \includegraphics[height=10.5cm]{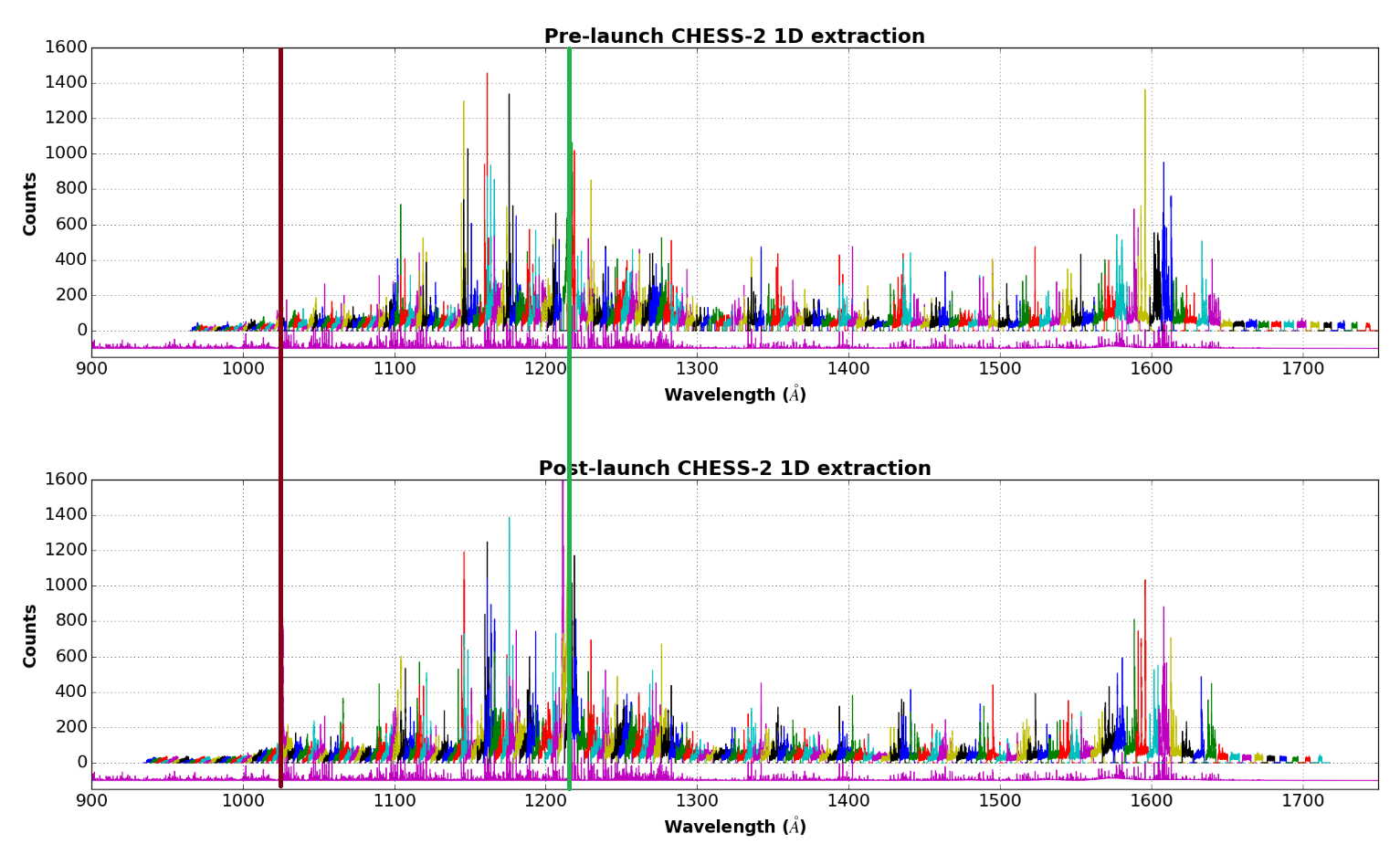}
   \end{center}
   \caption[fig12] 
   { \label{fig:fig12} 
Complete first-order wavelength solution for the pre- and post-flight CHESS-2 calibration spectra from 900 $-$ 1750 {\AA}. The final wavelength solution was determined using H$_2$ fluorescence emission features and a functional extrapolation of the wavelength with a 6$^{th}$-order polynomial fit, determined from pixel-to-wavelength solutions manually fitted for $\sim$30 orders with prominent emission features. Over-plotted in magenta is the model H$_2$ fluorescence inside the arc lamp (T$_{eff}$ = 800 K, N(H$_{2}$) $\sim$ 10$^{19}$ cm$^{-2}$, E$_{elec}$ = 50 eV). Each spectrum is scaled to the highest total counts of the H$_2$ features; otherwise, Ly$\alpha$ would dominate the spectrum and the H$_2$ features would be washed out in the images. To show how neighboring order spectra overlap and correlate to form the CHESS-2 1D spectrum, individual order spectra have been plotted using different colors. Overlaid are two vertical lines to show where Ly$\beta$ (dark red) and Ly$\alpha$ (green) are located in the spectra.
}
   \end{figure}

Once wavelength solutions were known for 20 $-$ 30 orders, we fit a 6$^{th}$ order polynomial function to extrapolate the wavelength calibration over the entire 130 orders in the CHESS-2 echellogram. Figure~\ref{fig:fig12} shows the final wavelength calibration for pre- and post-launch laboratory echellograms. The different colored spectra represent separate orders extracted from the echellograms. Because of the lower line density and AOI of the Bach echelle grating, we used higher-order dispersion solutions than were designed for CHESS. This resulted in overlapping wavelength solutions in adjacent orders, making it easier to stitch together the entire CHESS-2 1D spectrum from all order spectra via correlated spectral features. Despite changes in the echellogram between pre- and post-launch calibration images, there were no significant changes in the functional wavelength solution of the 1D extracted spectra. The only (minor) change to the functional wavelength solution was the shift of the starting pixel-to-wavelength conversion, since more shorter wavelength lines were available for extraction in the post-launch echellogram configuration. The final wavelength solution of the pre- and post-flight calibration spectra will provide more accurate radial velocity estimates of photospheric/ISM material and line identifications for atomic, ionized, and molecular material in the line of sight to $\epsilon$ Per.

%

%
   \begin{figure} [ht]
   \begin{center}
   \includegraphics[height=4.9cm]{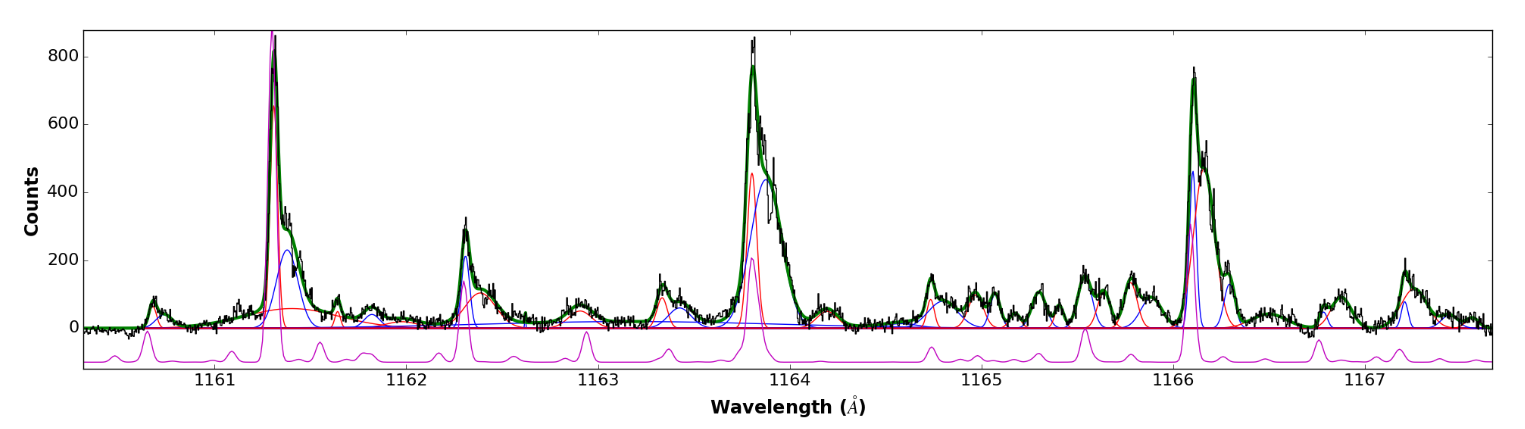}
   \end{center}
   \caption[fig13] 
   { \label{fig:fig13} 
LSF fits of H$_2$ emission features in the pre-launch calibration spectrum of CHESS-2 (echelle order m = 286, also featured in Figure~\ref{fig:fig11}). The order spectrum is shown in black. Red and blue Gaussian line fits are shown for the narrow and board Gaussian fits for each line, respectively. The green line is the sum of all Gaussian components to reproduce the spectrum. A modeled H$_2$ fluorescence spectrum is shown in magenta. We show how the power placed in the narrow and broad Gaussian components of each emission feature changes over the order spectrum; at shorter wavelengths, the narrow and broad Gaussian components hold similar percentages (50\%) of the power in the line, but as the wavelength increases across the order, more power is found in the broad Gaussian component (25\%/75\% in the narrow/broad components). This behavior is seen in order spectra with strong H$_2$ features, indicating that the cross disperser may be tilted and focusing shorter wavelength features in each order better than the longer wavelength features. For the three emission features, narrow line velocity widths vary between 11.8, 15.2, and 10.7 km/s respectively, while broad line fits have velocity widths between 34.1, 48.8, and 33.0 km/s, which are offset from the narrow line peaks by 17.9, 18.5, and 15.0 km/s.
}
   \end{figure}

After the wavelength solution was found for the pre-flight calibration echellogram, we determined the LSF and resolving power of the spectrum produced by CHESS-2. To determine the LSF of the instrument, we created a multi-Gaussian fitting routine to describe the line shapes of the emission features present in each echellogram order. An example of how this routine was applied for all emission lines across a given order is shown in Figure~\ref{fig:fig13}. At first glance, the emission lines produced by CHESS-2 are not symmetric; they have a sharp peak towards the left (blue-ward) side of the emission line, and a shallower slope back to the continuum level toward the right (red-ward). We described this line shape with two separate Gaussian functions summed together, which resulted in a narrow- and broad-component to each LSF. Each echelle order spectrum showed very similar behavior in LSFs, specifically that the shorter (longer) wavelength end of each order has more power in the narrow (broad) Gaussian fit. We see that the Area(narrow)/Area(broad) of the emission features for shorter wavelengths in the echelle order is $\sim$ 1, whereas emission features with longer wavelengths in the same echelle order have Area(narrow)/Area(broad) $\sim$ 0.33. This indicates that the cross disperser may be tilted, resulting in a better focus at one end of the echellogram. We note that for all emission lines, a peak (narrow) line fit was apparent, and we use this component of the LSF to estimate the resolving power of CHESS-2. We also consistently measured narrow line component velocity widths ($v_{narrow}$) in the pre-launch CHESS-2 spectrum to be between 3 $-$ 20 km/s, and broad line velocity widths ($v_{broad}$) to range between 20 $-$ 60 km/s. For post-flight calibrations, we measured $v_{narrow}$ $\sim$ 5 $-$ 25 km/s and $v_{broad}$ $\sim$ 20 $-$ 60 km/s.
All high S/N ($>$ 100) emission LSFs across the CHESS-2 bandpass were saved to later convolve with CHESS-2 flight data to eliminate instrumental effects seen in the final science spectrum.

%
   \begin{figure} [ht]
   \begin{center}
   \includegraphics[height=7cm]{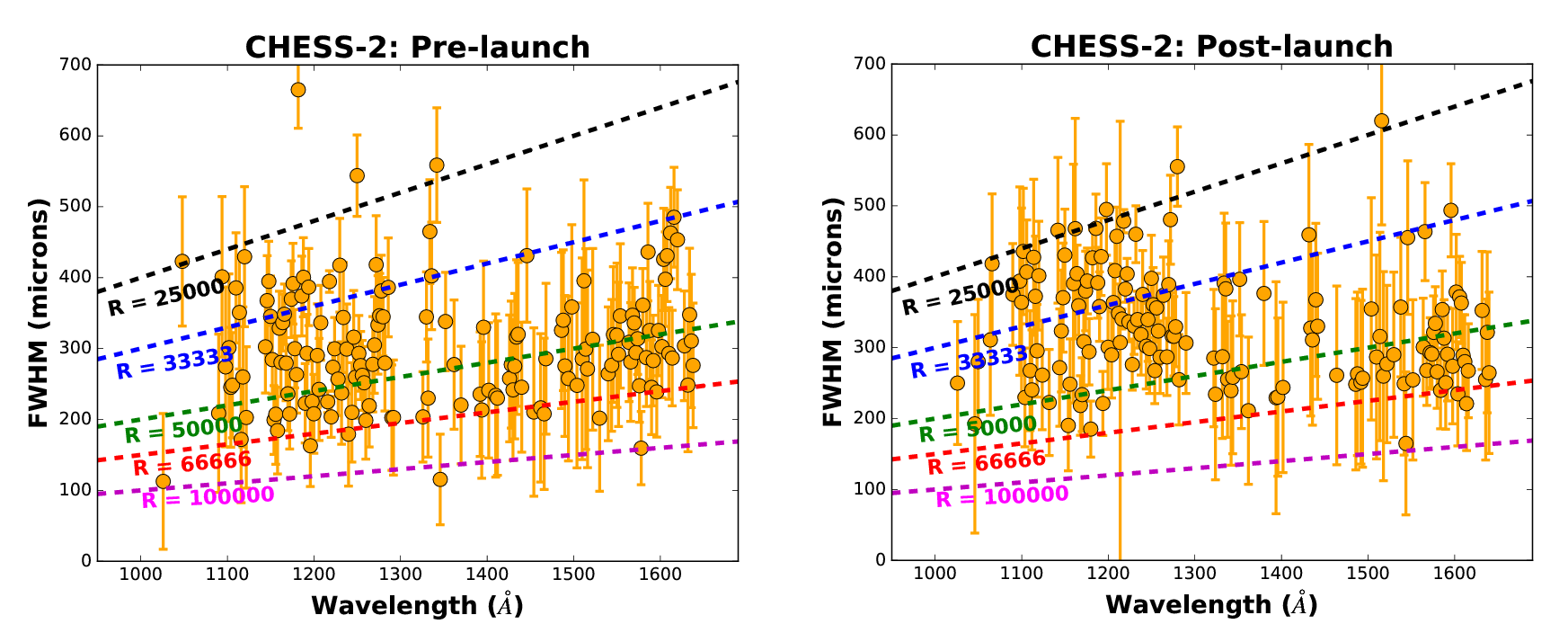}
   \end{center}
   \caption[fig14] 
   { \label{fig:fig14} 
A scatter plot of the measured resolution (via FWHM measurements) of individual electron-impact H$_2$ emission features in the pre-flight calibration spectrum of CHESS. Orange spots are measured FWHM values of individual line peak cores, and the dashed lines represent resolution cutoffs, based on the FWHM (in microns) of the emission line as a function of wavelength. Both pre- and post-launch resolving powers seem to be concentrated around 25,000 $-$ 70,000 over the bandpass of CHESS-2, but there is a noticeable concentration of lines at lower resolving powers (larger FWHM) at $\lambda$ $<$ 1300 {\AA} for the post-launch calibration. Physical shifts in the optical alignment between pre- and post-launch calibrations affected the resolving power of the instrument. This may have been caused by tilting the cross disperser and subsequent echellogram, or by moving away from the instrument focus, or a combination of both. 
}
   \end{figure}

We show the FWHM of the narrow component LSFs of the emission lines observed across the CHESS bandpass in Figure~\ref{fig:fig14}. Each orange spot represents a different emission line in the CHESS-2 calibration image, and the different dashed lines show the resulting resolving power as a function of FWHM and wavelength (black: R = 25,000; blue: R = 33,333; green: R = 50,000; red: R = 66,666; magenta: R = 100,000). The pre-launch calibration shows a concentration of resolving powers between R = 25,000 and 66,000, with an average resolution between R $\sim$ 33,000 $-$ 70,000 (velocity width between 9.0 $-$ 4.5 km/s) across the CHESS-2 bandpass. While this is below our nominal resolving power goal of 100,000, we note that the measured resolving power of CHESS-2 using the laboratory calibration data may only represent a lower limit to the instrument capabilities. As mentioned previously by Ref.~\citenum{Hoadley+14}, the ray trace for CHESS requires an input on-axis light source with beam spread $<$ 1$''$, but we have only been able to demonstrate constraining the spread to 2$''$ $-$ 3$''$ in our laboratory vacuum chamber. Additionally, the laboratory arc lamp may have pressure-broadening effects on the H$_2$ electron-impact emission. This would produce emission lines with FWHMs larger than the resolving power of the instrument, resulting in only being able to measure a lower limit to the resolving power of CHESS-2 across the bandpass. 

However, post-launch calibration results show a shift from the average R $\sim$ 33,000 $-$ 70,000 to R $\sim$ 25,000 $-$ 60,000 (velocity width between 12.0 $-$ 5.0 km/s). This indicates that either the instrument shifted slightly out of focus before and/or during launch operations, or the shift in the echellogram affected the line shapes of the spectra. In either case, post-launch resolution of CHESS-2 is degraded from pre-launch values. New LSFs were measured for the post-launch CHESS-2 calibration echellogram to better correct for instrument affects in the science spectrum.

\section{CHESS-2 Launch and Preliminary Flight Results}
\label{sec:flight}

CHESS-2 was brought to White Sands Missile Range (WSMR) in late January 2016 for field operations in preparation for launch. CHESS-2 underwent various tests, including vibration, which required a means of determining alignment shifts before launch. We fitted a Bayard-Alpert tube (ionization gauge)\cite{McCandliss+00} with a small, collimating mirror and pinhole (20 $\mu$m) to the shutter door, which produced an echellogram with air spectral features (C, N, O, H). We used these images to measure the centroid pixel location (in both x and y axes) of N I and O I emission features. Comparing the new centroid location of 3 $-$ 4 emission lines to reference pixel locations measured at CU, we use the plate scale of the instrument (plate scale = 206265$''$/1236.834 mm = 166.77 $''$/mm) to estimate alignment shifts before launch. The largest centroid shift measured pre-launch was 105 pixels (over 2k $\times$ 2k pixels), which corresponded to a physical shift of 1176 $\mu$m (1.176 mm), or 3.27$'$ (196.12$''$). Alignment shifts are apparent between pre- and post-launch calibration images (Figure~\ref{fig:fig10}). Given the large FOV of the instrument (0.67$^{\circ}$, or 40.2$'$), the success criteria specification for the on-target acquisition (5$'$), and the ability to demonstrate that the instrument can still collect a science echellogram, this alignment shift was acceptable to continue with launch of the instrument.

CHESS-2 was launched aboard NASA mission 36.297 UG from White Sands Missile Range (WSMR) on 21 February 2016 at 09:15pm MST using a two-stage Terrier/Black Brant IX vehicle. The mission was deemed a comprehensive success. The instrument successfully collected data over the allotted $\sim$ 400 seconds of observing time, with $>$ 200 seconds without up-link maneuvers. When the instrument centered on-target ($\epsilon$ Per), the count rate was lower than expected ($\sim$ 25,000 counts/second, instead of 50,000 counts/second), and as much as half of the signal was geo-coronal scattered light. We steered the payload around the FOV for the first half of the flight, to see if we were off-axis from the star, but we elected to move back to the original acquisition position and integrate for the second half of the flight.

%
   \begin{figure} [ht]
   \begin{center}
   \includegraphics[height=7.3cm]{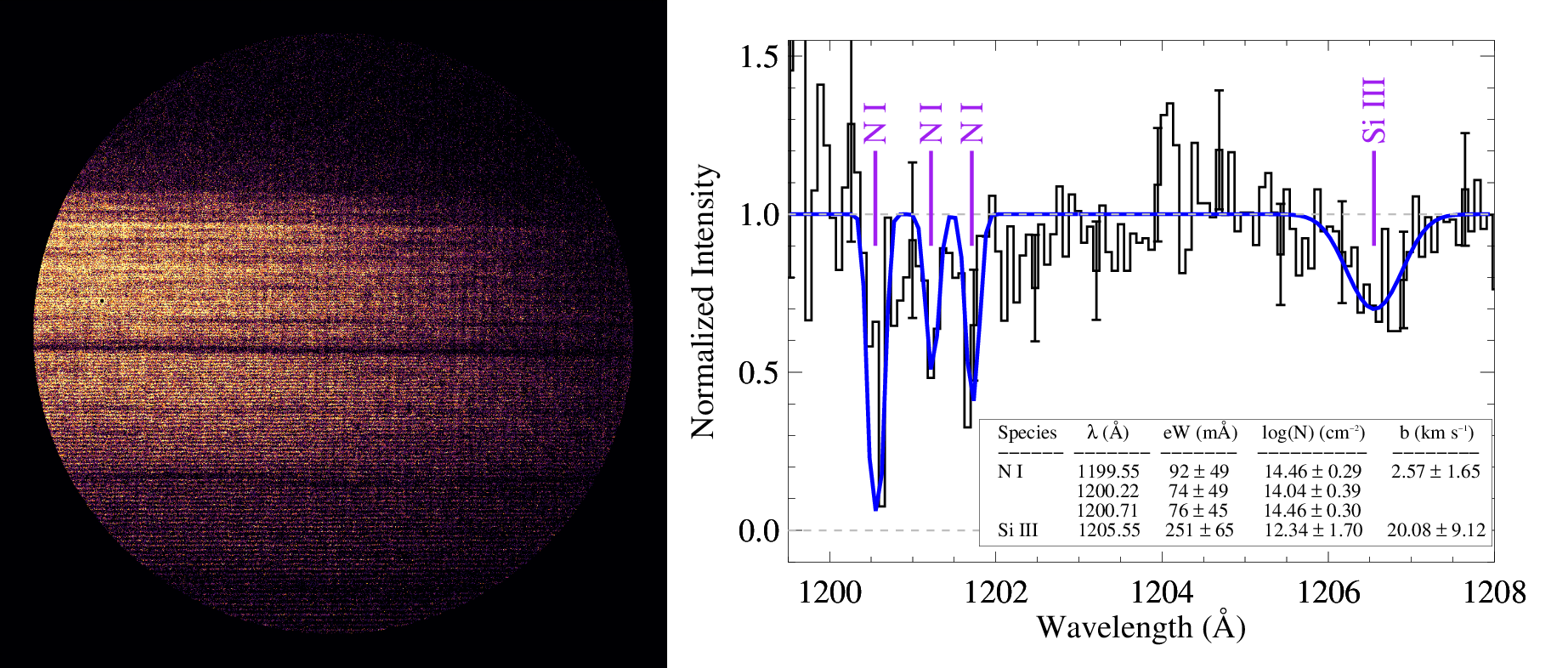}
   \end{center}
   \caption[fig15] 
   { \label{fig:fig15} 
The flight data from 36.297 UG (CHESS-2). \emph{Left:} The science echellogram of $\epsilon$ Per after an exposure time of $\sim$ 200 seconds, with inter-order scatter subtracted from the raw image. The echelle orders are stacked horizontally in the image, with order spectra easily distinguishable in the bottom half of the echellogram. Because the echelle used in CHESS-2 disperses the starlight into very high orders for $\lambda$ $<$ 1200 {\AA} (m $>$ 280), shorter wavelength orders are more difficult to distinguish and required scattered light subtraction and echellogram collapsing along the order axis of the image. Progress is still being made to better identify and extract the spectra from these orders. \emph{Right:} A normalized spectrum extracted from one order identified in the raw fight data to show interstellar absorption features in the data. The order (m = 276) ranges from $\sim$1199 $-$ 1208 {\AA} and shows warm (Si III) and cool (N I) interstellar features against the stellar continuum. Preliminary line fits are shown in blue, and physical quantities derived from the line fits are listed in the bottom right of the plot. Both N I and Si III column densities and b-values are consistent with line of sight cool and warm ISM diagnostics previously explored in the literature\cite{Redfield+04,Snow+77}.
}
   \end{figure}

From the beginning of the flight, we immediately saw photospheric and interstellar absorption features in the echellogram of $\epsilon$ Per, the prominent features being Ly$\alpha$, O I, and C III, as well as stellar continuum for orders with $\lambda$ $>$ 1300 {\AA}. Once we integrated on $\epsilon$ Per during the second half of the flight, interstellar features started to appear, including Si III, N I, Si II, and H$_2$ complexes. Figure~\ref{fig:fig15} shows the raw flight data (echellogram; left) after $\sim$ 200 seconds on-target, including an extraction of one order that shows both the N I and Si III absorption features (right). After post-processing of the data (see Section~\ref{sec:post}), a preliminary analysis of the N I and Si III absorption features has been completed. Our analysis assumed a Gaussian line profile LSF convolved with an instrument resolving power R = 50,000. The results of the N I and Si III abundances and b-values, which represents both the turbulent velocity and temperature ISM species, are consistent with cool and warm interstellar diagnostics explored in the local ISM\cite{Redfield+04,Snow+77}. 

Overall, the quality of the flight was S/N $\gtrsim$ 5, which is sufficient to identify spectral absorption features, but may make it difficult to quantify the column densities and b-values of the line-of-sight measurement. We are currently working on techniques to better extract the stellar spectrum from the geo-coronal background (discussed in Section~\ref{sec:post}), making it easier to identify atomic and molecular contents in the science data.

\section{Future Work and Launches}
\label{sec:future}

\subsection{Post-Flight Calibration and Science Data Reduction}
\label{sec:post}

We verified that all essential components of the instrument were working and completed post-launch activities for CHESS-2 upon return from WSMR after the launch of 36.297 UG. First, because we elected to stay on $\epsilon$ Per for the entirety of the rocket flight, we needed an understanding of how the geo-coronal scattered light may spread across the detector before beginning post-launch calibrations. Therefore, we took a scattered light spectrum with the field operations Bayard-Alpert tube fitted to a mini-Conflat opening in the instrument shutter door. This spectrum provided a rough ``flat-field" tool to apply to the science echellogram. Then, we integrated the instrument into our vacuum chamber and took a deep H/Ar spectrum, complementary to the one taken pre-launch. Post-launch calibrations are described with pre-launch calibrations in Section~\ref{sec:cals}. One of the biggest surprises during 36.297 UG was the lower-than-expected count rate while on $\epsilon$ Per. One possibility is that the efficiencies of the gratings and detector were over-estimated before instrument build-up or degraded between instrument build-up and launch. The former seems more plausible, given that the pre- and post-launch calibration exposures were comparable for similar arc lamp conditions. Another possibility is that we aligned the echellogram to the wrong echelle orders. For example, instead of aligning to the main and next brightest adjacent order for a total of 20\% efficiency (at Ly$\alpha$), we aligned the echellogram to the main order and less bright adjacent order, giving a lower$-$than$-$expected total order efficiency (likely closer to 10\% at Ly$\alpha$). In either case, we will be vigilant in the next iterations of CHESS to be certain we align our instrument properly and keep our optical coating pristine during instrument build-up and alignments.

%
   \begin{figure} [ht]
   \begin{center}
   \includegraphics[height=5.1cm]{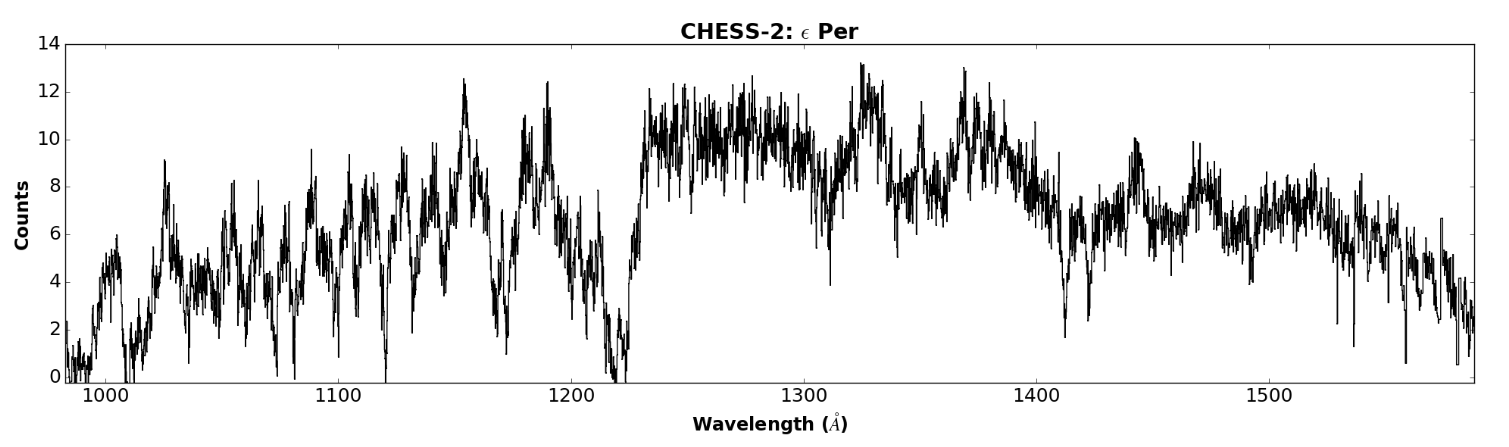}
   \end{center}
   \caption[fig16] 
   { \label{fig:fig16} 
The full 1D spectrum of $\epsilon$ Per, with a preliminary wavelength fit, from the $\sim$ 200 seconds of data taken on 36.297 UG. Order spectra have been cross-correlated, with photon counts summed to create a 1D spectrum spanning the CHESS bandpass (1000 $-$ 1600 {\AA}). Spectral features are apparent in the data, including the broad Ly$\alpha$ feature (1215.67 {\AA}), Si III (1206.51 {\AA}), C III (1174.26 $-$ 1176.37 {\AA}, appears blended), and several H$_2$ complexes ($\sim$ 1020 {\AA}, 1050 {\AA}, and 1070 {\AA}). Continued effort is being made to separate the geo-coronal background from the stellar continuum via extracting scattered light profiles from ``dark'' regions of the science data and subtracting inter-order scatter from the echelle orders, which will help with order extraction, cross-correlation of order spectra, and refinement the wavelength solution of the science spectrum.
}
   \end{figure}

Figure~\ref{fig:fig16} shows a preliminary extraction of the CHESS-2 science data, with a preliminary wavelength solution determined from post-launch calibrations. Post-processing of the science data is on-going, with a focus on subtracting the geo-coronal background from the stellar spectrum. Elimination of the scattered light profile in the flight data allows for better extraction of the echelle order locations in the echellogram, cross-correlation of absorption features in the science spectrum, and better definition of the wavelength solution for the final science spectrum. One way to eliminate the scattered light is to estimate the noise in the data via extraction of photon counts in echellogram ``dark'' regions $-$ either regions on the detector where the instrument effective area is very small ($\lambda$ $<$ 900 {\AA}) or absorption lines we know to be saturated, such as Ly$\alpha$ and O I - and subtracting the scattered light profiles in these regions from similar areas in the echellogram. The geo-coronal background can also be estimated from the inter-order contamination in the flight echellogram, which is determined from taking the sum of all photon counts along the x-axis and extracting the noise from the minima pixel values between echelle orders. This technique minimizes the inter-order counts and exaggerates order spectra locations in the echellogram, making the stellar continuum-to-absorption line ratios deeper in the 1D extracted spectra.

\subsection{Future Launch and Instrument Changes}

The CHESS instrument is scheduled to launch two more times $-$ CHESS-3 (36.323 UG) is currently set to launch in June 2017 from WSMR, and CHESS-4 will launch in early 2018, ideally during an Australian sounding rocket campaign. CHESS-3 will implement a new echelle from Richardson Grating (87 grooves/mm, 63$^{\circ}$ blaze angle), which has demonstrated $>$ 50\% peak order efficiency at Ly$\alpha$ (see Figure~\ref{fig:fig6}) and has order solutions identical to those designed for the CHESS instrument (m = 266 $-$ 166). CHESS-4 will see the implementation of an alternative detector technology, the $\delta$-doped charge coupled device (CCD), with new readout technologies to accommodate large focal plane arrays\cite{Veach+13}.

\acknowledgments 
 
The authors would like to thank the students and staff at CU for their tremendous help in seeing CHESS-2 come to fruition. We would also like to thank the NSROC staff at WFF and WSMR for their tireless efforts that pushed us to a smooth launch. NK also thanks Rocket League and the song ``Sister Christian" by Night Ranger for keeping him sane through field operations in February 2016. This work was supported by NASA grant NNX13AF55G. 

\bibliography{references} 
\bibliographystyle{spiebib} 

\end{document}